\documentclass[12pt,german]{article}
\usepackage{amssymb}
\usepackage{epsfig}
\begin{document}
\title{\bf Quantum Zeno Features \\ of Bistable Perception \vskip 1.5cm }
\author{{\bf Harald Atmanspacher} \\[0.1cm]
Institut f\"ur Grenzgebiete der Psychologie und Psychohygiene\\
Wilhelmstr.~3a, D--79098 Freiburg; \\
Max-Planck-Institut f\"ur extraterrestrische Physik \\
D--85740 Garching \\[0.2cm]
\and 
{\bf Thomas Filk and Hartmann R\"omer} \\[0.1cm]
Institut f\"ur Physik, Universit\"at Freiburg \\
Hermann-Herder-Str.~3, D--79104 Freiburg }

\date{}

\maketitle

\vskip 2cm
\begin{abstract}

A generalized quantum theoretical framework, not restricted to the validity
domain of standard quantum physics, is used to model the dynamics 
of the bistable perception of ambiguous visual stimuli. The central idea 
is to treat the perception process in terms of the evolution of an unstable
two-state quantum system, yielding a quantum Zeno type of effect. A quantitative
relation between the involved time scales is theoretically derived. This relation is found 
to be satisfied by empirically obtained cognitive time scales relevant for bistable perception.

\end{abstract}
\newpage

\section{Introduction}

Quantum theory has revolutionized our understanding of 
the physical world in both scientific and epistemological respects. 
It was developed in the third decade
of the $20^{\rm th}$ century as a theory describing the 
behavior of atomic systems. Subsequently, its range of validity turned
out to be much wider. Not only are nuclei and elementary particles, more 
than seven orders of magnitude smaller than atomic systems, 
governed by quantum theory, but also macroscopic phenomena like
superconductivity or superfluidity are successfully 
described in quantum theoretical terms.


>From the present state of knowledge, the broad validity of quantum theory in physics is
not surprising. Investigations of its conceptual structure and axiomatic foundations have 
revealed that quantum theory is a logical consequence of some rather simple 
and plausible basic assumptions on the nature of observables and states of 
physical systems. In this framework, classical physics results from basically
one additional assumption: the commutativity of the algebra of observables. 
On the other hand, the axiomatic framework has shown how deeply rooted 
apparently bizarre concepts of quantum theory like complementarity and
entanglement really are, and that there is no way back to classical concepts on
a fundamental level of physical description.

Since the early days of quantum theory, starting with Niels Bohr, the idea has 
been entertained that quantum theoretical
concepts like complementarity and entanglement might be meaningful
and important even beyond the realm of the physical world in its
strict sense. Concerning such ``quantum-like'' phenomena, encorporating 
complementarity or entanglement beyond physics, three different stances, not
completely exclusive with respect to each other from a logical point of view, 
are possible and have found proponents.
\begin{enumerate}
\item
The application of quantum theoretical concepts to non-physical situations
is merely metaphoric and not elucidated in a precise and general
way. The analogies are selected ad hoc and change with the given situation.
\item
In the spirit of a strict physicalism, everything can be reduced to physics. 
There is nothing left beyond the world of physics, and quantum-like phenomena 
as adressed above are simply physical quantum phenomena.
\item
Assuming a non-reductive framework, it might be possible and adequate to 
describe vital features of parts of the world, physical or non-physical, 
by a formalism isomorphic to the formalism of quantum theory. This isomorphy may 
be total or, more often, partial, such that only particular features of the 
quantum theoretical formalism are realized in a non-physical context.
\end{enumerate}

An approach within option (3), denoted as weak quantum theory, was recently proposed on 
an axiomatic basis \cite{atm1}. The starting point in this approach is the algebraic formulation 
of quantum theory, where the observables of a physical system generate a $C^*$-algebra
$\cal{A}$, and the states of the system are positive linear functionals on
$\cal{A}$. A careful analysis of this formulation shows that its axioms are not
equally fundamental. They can be sifted out into general, indispensable axioms, 
applying to any (physical or non-physical)
system as long as it is a possible object of meaningful investigation, and more 
special axioms used in quantum physical applications only.

The most general formal system thus arising, in which concepts like
complementarity and entanglement are still meaningful, is a minimal version of
weak quantum theory. In short, it can be described in the following way
(for details we refer to \cite{atm1}):

The notions of system, observables, and states remain unchanged. Observables are 
identified with functions mapping states to states. This emphasizes the importance of
observables as representing active processes with the capacity of changing states, 
which is vital in quantum theory.

Observables, conceived as functions, can be composed. This composition defines a
multiplication of observables endowing the set of all observables with the
simple structure of a semigroup. Two observables are incompatible, or complementary 
with respect to each other, whenever they do not commute. Entanglement can be 
rephrased as complementarity of a global observable pertaining to a system as a whole
and local observables pertaining to its parts. Propositions are special observables
related to yes-no-questions about the system.

The minimal, most general formal system of weak quantum theory can be stepwise
supplemented and enriched up to isomorphy with the full quantum theory used in physics.
In its minimal form, weak quantum theory, although it implies complementarity
and entanglement, is vastly more general and flexible than the full quantum
theory. The main differences are:
\begin{itemize}
\item
In the most general form of weak quantum theory, the sum of observables is
not defined in an empirically meaningful way. As a consequence, there is no 
Hilbert space of states and no way to attribute
probabilities to the outcomes of measurements corresponding to the observables.
\item
Planck's constant $h$, which determines the amount of non-commutativity
and complementarity in quantum physics, enters nowhere. This implies
that conspicuous quantum-like phenomena may be effective in situations beyond
standard quantum physics.
\item
Unlike in quantum theory, where Bell's inequalities can be derived, there
is in general no way to rule out local realism. As a consequence, indeterminacies 
and complementarities can be of epistemic rather than ontic origin.
\end{itemize}

Quantum-like behavior as described by weak quantum theory or its
refinements is expected for systems with complex internal organization. 
This implies strong and intricate coupling of their parts and the practical 
impossibility of observation from outside without influencing the state of such 
systems.

A paradigmatic example for such a situation is the cognitive system and its
neural correlates in the brain. This may be one of the reasons why quite a
number of approaches have been developed, in more or less detail, to relate
conscious activities to quantum theory or describe particular brain operations
quantum theoretically. Besides the fairly popular approaches associated with 
the names of von Neumann/Wigner/Stapp \cite{stap1} and Penrose/Hameroff \cite{pen},
other interesting options have been proposed by Beck/Eccles \cite{beck} and 
Umezawa/Vitiello \cite{umez}. 

All these approaches are essentially discussed within the usual quantum physics. 
This is at variance with the strategy of weak quantum theory; a recently discussed example 
refers to the perception of temporally subsequent events \cite{atm2}. This approach
proposes a type of temporal entanglement which is not contained in standard quantum
physics, but can be formulated in the theory of chaotic systems. In the present
paper, we intend to apply the framework of weak quantum theory to another 
scenario of cognitive science: the bistable perception of ambiguous 
stimuli.\footnote{Niels Bohr was familiar with the bistable perception 
of ambiguous stimuli through his psychologist friend Edgar Rubin. There are 
indications \cite{plau} that this fact, together with Bohr's studies of the
writings of Harald H{\o}ffding and William James, played an important role 
in the complicated genesis of his concept of complementarity in quantum physics.}

Bistable perception arises whenever a stimulus can be interpreted
in two different ways with approximately equal plausibility. 
A very simple and often investigated example of bistable perception 
is the so-called Necker cube. (For an overview
concerning the current discussion of cognitive and neural features of the perception
of the Necker cube see \cite{korn1,korn2}.) A grid of a cube in
two-dimensional representation can be perceived as a three dimensional-object in 
two different perspectives, either as a cube seen from above or from below. 
The perception of the Necker cube switches back and forth between
the two possible interpretations spontaneously and inevitably.

We propose to describe bistable perception with the formalism of a 
two-state quantum system, where the two basis states correspond to the two
different ways to interpret the visual stimulus. Measurement is considered as
the mental process determining in which way the figure is perceived.
The switching between the different perceptions corresponds to the quantum
transition between the two states which are eigenstates of the operator 
representing a particular perception and unstable under the time evolution
of the system.

Such a description of bistable perception employs a non-minimal version
of weak quantum theory with a linear structure and a two-dimensional
linear state space. This version is fairly close to the structure 
of the full quantum theory used in physics. This does not imply, however,
that we propose to understand bistable perception as a quantum 
phenomenon in the sense that the related brain processes are usual quantum
processes. (Planck's constant $h$ will nowhere enter in our arguments.)
Rather, we will discuss the quantum-like behavior of bistable perception
as a result of the truncation of an extremely complicated system 
to a two-state system, into which the effect of many uncontrolled variables 
and influences is lumped in a global way.
 
For this purpose, we
consider a quantum mechanical system with a state space spanned by
two states $\psi_1$ and $\psi_2$, neither of which is an eigenstate of the
Hamiltonian $H$ generating the evolution matrix $U(t)={\rm e}^{{\rm i}Ht}$. 
If the system is initially in the state $\psi_1$ and
allowed to evolve freely according to $U(t)$, then its state will oscillate
between the states $\psi_1$ and $\psi_2$. This oscillation can be  
slowed down by increasing the frequency at which the system is measured, 
asking whether it still resides in its initial state. In the limit 
of continuous measurement, the evolution of the system can be completely 
suppressed. This phenomenon is known as the quantum Zeno 
effect.\footnote{See \cite{home} for a review of theoretical and
experimental results concerning the quantum Zeno effect. Its possible 
cognitive significance was indicated previously by Stapp \cite{stap2}.}

In the following section 2, the quantum Zeno effect will be described
in as much detail as required for the purpose of this paper. The
quantitative relation between different time scales of crucial significance
will be emphasized in particular. In section 3, cognitive time scales satisfying this
relation will be presented with particular respect to the dynamics of bistable perception. 
Section 4 summarizes the results and concludes the article. 

\section{Quantum Zeno effect}

The quantum Zeno effect was originally discussed as the quantum Zeno 
``paradox''\footnote{The Greek philosopher Zeno of Elea proposed the following
antinomy: ``As long as anything is in space equal to itself, it is at rest.
An arrow is in a space equal to itself at every moment in its flight, and
therefore also during the whole of its flight. Thus the flying arrow is at 
rest.'' \cite{cajo}} 
by Misra and Sudarshan \cite{misr} for the decay of unstable quantum systems.
As mentioned above, its key meaning is that repeated observations of the
system decelerate the time evolution which it would undergo without observations,
e.g.~its decay. The metaphor ``a watched pot never boils'' paraphrases this
behavior in the limit of continuous observation. 

The situation addressed in the following refers to a quantum system oscillating 
between two non-stationary states.
For this purpose, we consider a system with the following properties:
\begin{enumerate}
\item   
For convenience, a two-state system will be considered. (The results apply to 
more general systems as well.)
\item   
An observation is represented by the operator
\[    \sigma_3 = \left( \begin{array}{cc}
        1 & 0 \\ 0 & -1 \end{array} \right) . \]
Immediately after an observation, the system will be
in one of the corresponding eigenstates
\[  \psi_1 =  |+ \rangle = \left( \begin{array}{c}
     1 \\ 0 \end{array} \right)
   \hspace{1cm} {\rm or} \hspace{1cm}
    \psi_2 =  |- \rangle = \left( \begin{array}{c}
     0 \\ 1 \end{array} \right) . \]
\item  
Both $\sigma_3$-eigenstates may also be represented
by their projection operators
\[  P_+ = \left( \begin{array}{cc}
        1 & 0 \\ 0 & 0 \end{array} \right)
   \hspace{1cm} {\rm and} \hspace{1cm}
    P_- = \left( \begin{array}{cc}
        0 & 0 \\ 0 & 1 \end{array} \right) . \]
\item  
Without loss of generality, the Hamilton operator giving rise to transitions of the 
system can be written as 
\[     H = g \sigma_1 = g
      \left( \begin{array}{cc}
        0 & 1 \\ 1 & 0 \end{array} \right) , \]
where $g$ is a coupling constant. Hence, 
the unitary operator of time evolution is represented by
\[  U(t) = {\rm e}^{{\rm i}Ht} =
   \left( \begin{array}{cc}
      \cos gt & {\rm i} \sin gt             \\
     {\rm i} \sin gt   & \cos gt  \end{array} \right). \]
\item  
In this model, $\Delta T$ defines the time interval between
two successive observations, and $T$ defines the time scale
after which the state has changed with $50\%$ probability.
Concerning the cognitive interpretation of
$\Delta T$ and $T$ we refer to the next section. It is assumed that
$T/\Delta T =N \gg 1$.
\end{enumerate}

We now calculate the probability that an eigenstate of
the observation operator  $\sigma_3$
(representing the perception of the Necker cube)
remains unchanged after a time $T=N\Delta T$ under the
condition that repeated observations (measurements of
$\sigma_3$) occur in time intervalls $\Delta T$.
For $t=0$ we assume the system to be in the eigenstate
$|+\rangle$, and this state is confirmed after each
observation.

The probability that the system is still in state $|+\rangle$ after time $t$ is:
\begin{equation}
\label{cosgt}
w(t) =
  \left| \langle + | U(t) | + \rangle \right|^2
   =     \cos^2 gt \, .
\end{equation}
This oscillation determines a characteristic time scale
$t_0$, given by the requirement that the state of
the system contains the eigenstates $|+\rangle$ and
$|-\rangle$ with equal probability, corresponding to
$1/8$ of a period of the oscillation or $t_0=\pi/4g$.

By contrast, if the system is observed $N$ times with time step
$\Delta T$, the probability that it is still in state $|+ \rangle$
after all $N$ observations, each with the result $+$, is given by
\begin{equation}   w(N) =
  \left| \langle + | (P_+ U(\Delta T) P_+)^N
             | + \rangle \right|^2
   =     \left[ \cos (g\,\Delta T) \right]^{2N} \, .
\end{equation}
This is simply the product of the survival probabilities for each 
individual observation. The number of observations $N$ after which 
the probability $w(N)$ decreases to $1/2$ is given by
\begin{equation}
\label{wN}
   w(N) = \frac{1}{2} ~~ \rightarrow ~~
   [\cos (g\, \Delta T)]^{2N} = \frac{1}{2}  \, ,
\end{equation}
or
\[  \cos (g\,\Delta T) = {\rm e}^{-\frac{1}{2N} \ln 2} \, . \]
According to $N=T/\Delta T \gg 1$, the right hand side of this equation is close
to unity. Therefore, the argument of the cosine is
close to zero and we may expand the cosine function according to
\begin{equation}
\label{exp}
  \left( 1 - \frac{(g\, \Delta T)^2}{2} + \ldots \right)
   \approx 1 - \frac{1}{2N} \ln 2  + \ldots
\end{equation}
or
\begin{equation}
\label{g}
     g = \sqrt{\frac{\ln 2}{N \, \Delta T^2}}  =
          \sqrt{\frac{\ln 2}{T \, \Delta T}}\, .
\end{equation}
In this way, the unknown coupling constant $g$ is expressed
by the experimentally accessible quantities $\Delta T$ and $T$.

These quantities can be related to the evolution of the
system under the condition that no observations are performed. In this case,
the time evolution is given by $U(t)$ and, as mentioned above, 
the state oscillates between
the two eigenstates $|+\rangle$ and $|-\rangle$ with period $t_0 = \pi/4g$. 
This leads to the relation
\begin{equation}
\label{ergebnis}
  t_0 =
   \frac{\pi}{4 \sqrt{\ln 2}} \sqrt{T\, \Delta T}\, .
\end{equation}
between the three time scales involved. (Note that Planck's action $h$ is
absent in this relation. If at all, it would enter in $g$, but $g$ is eliminated in
eq.~\ref{ergebnis}.)  

The derivation of this relation depends on two
arbitrary choices: $T$ is determined from the condition that the
probability of state flipping is $1/2$ (eq.\
\ref{wN}) , and $t_0$ is determined from the condition that
the oscillating state is a superposition of
eigenstates of $\sigma_3$ with equal
coefficients (eq.~\ref{ergebnis}). Even if these conditions are
varied, the general result
\begin{equation}
\label{allgemein}
      t_0 =  C \sqrt{ T \Delta T} \, ,
\end{equation}
remains unchanged ($C\approx 1$; cf.~eq.~\ref{ergebnis}). It entails the following two predictions:
\begin{enumerate}
\item
As long as the time interval $\Delta T$ between two
observations is non-zero, the states will spontaneously switch
into each other after an average time $T$, which is large compared 
to $\Delta T$ and $t_0$.
\item
The relation between the time scales $T$,
$\Delta T$ and $t_0$ is given by eq.~\ref{allgemein}.
\end{enumerate}

\section{Cognitive time scales}

In order to assign significance to the time scales $T$, $\Delta T$ and $t_0$ 
in terms of the process of bistable perception, corresponding cognitive time 
scales have to be identified with particular respect to bistable perception. 
In this section, we will argue that there are natural choices for $T$ and 
$\Delta T$. As a consequence, $t_0$ can be calculated, and its possible
significance will be discussed. 

\subsection{$T \approx 3$ sec}
   
The perception of ambiguous visual stimuli is a prominent topic of modern research in
cognitive science and neurophysiology \cite{kruse}. 
One of the elementary examples is a two-dimensional image of a three-dimensional cube, 
the so-called Necker cube. The Swiss geologist Necker \cite{neck} first discovered that 
the front-back orientation
of the cube switches spontaneously. Since then, numerous other stimuli have been studied
generating the same basic phenomenon. Due to its simplicity and, in comparison with other stimuli, 
fairly low semantic content, the Necker cube remains one of the most popular objects for
investigation.  

There are two basically different approaches to study the perception of ambiguous stimuli. The 
first one refers to the behavioral response to a stimulus which is (assumed to be) based
on psychological (mental) processes. Measuring the frequency of reversals is a
typical example. The second perspective is to look for neural correlates of psychological
processes triggered by stimuli, using either electrophysiological tools or, more recently,
imaging techniques.     

One of the fairly invariant patterns, which the perception of ambiguous stimuli
presents, is a remarkably stable rate of reversals for individual subjects, ranging between about 4 
and 60 switches per minute for different subjects \cite{brown}. This reversal rate 
corresponds to a ``mean first passage time'' between 1 and 15 seconds. The duration after 
which the stimulus orientation spontaneously reverses was found to be gamma-distributed 
around a maximum of about 3 seconds \cite{dema}. This time scale
can straightforwardly be attributed as the extended oscillation period $T$ 
due to observations.    

The time scale of approximately 3 seconds was not only found in the bistable perception
of visual stimuli, but also of auditory ambiguous stimuli. For instance, the phoneme sequence 
BA-CU-BA-CU-... switches into CU-BA-CU-BA-... after a corresponding time interval
\cite{radi} . 
In addition, there are other features in perception, cognition, memory, and movement 
control for which the 3 second interval is crucial. Here are some of the most conspicuous 
observations, discussed in more detailed in \cite{poep2}:
\begin{itemize}
\item length of lines in classical verse in different languages \cite{kien}, also of melody phrases;
\item segmentation of spontaneous speech acts, i.e.~closed verbal utterances \cite{kowa};
\item rhythmic accentuation of successive beats within 3 second windows \cite{szela};
\item length of spontaneous motor activity (e.g., scratching) in different mammalian species \cite{gers};
\item information retrieving by short-time (working) memory within 3 second windows \cite{pete}    
\item the reproduction of time intervals is overestimated for time intervals smaller than 3 seconds,
and underestimated for intervals greater than 3 seconds; the ``indifference point'' is at 3 seconds \cite{poep3}.
\end{itemize}  

All these observations and more indicate that temporal segmentation into 3 second windows is
a basic principle of many aspects of conscious activity. Therefore P\"oppel refers to single 3 second
intervals as ``states of being conscious'' and emphasizes  that
the segmentation mechanism itself is automatic and presemantic \cite{poep2}. Such discrete
successive states are semantically linked with their predecessor and successor. The resulting
subjective experience of the continuity of consciousness is, thus, intimately related to the 
assignment of meaning.    

The ubiquity and the basic significance of the time scale of approximately 3 seconds suggests
that $T \approx 3$ sec is also significant for cognitive processes beyond the bistable perception
of ambiguous stimuli. In the context of the present article, however, the focus remains bistable 
perception of Necker type stimuli, since it offers a scenario which is conceptually better defined and 
experimentally better controllable than transitions between arbitrary other mental representations.  

\subsection{$\Delta T \gtrapprox 30$ msec}

A reasonable estimate for the time between observations in the sense of the
quantum Zeno effect, $\Delta T$, is difficult to obtain from the phenomenology 
of bistable perception. It has to satisfy at least one condition: the
perceptual system must be able to assign a temporal sequence to successive
events, i.e.~observations.

Experiments concerning the capabilities of discriminating and
sequentializing temporally separate perceptual events have been carried out
for a long time \cite{stei}. 
A particular version of such experiments was reported by P\"oppel \cite{poep1}. 
Exposing subjects to two successive separable (e.g.~by frequency) stimuli 
and varying the time interval $\Delta t$ between them, three regimes of different 
kinds of perception of the stimuli were observed. 

For $\Delta t \gg 30$ msec, two different individual events are clearly 
separable, and their sequence can be correctly assigned. 
For $\Delta t < 3$ msec (in the auditory modality), the two different events 
remain unresolved and, as a 
consequence, a sequential order cannot be assigned to them. Most interesting 
is the result for the regime 3 msec $ < \Delta t < $ 30 msec. 
Here, two individual
different events can be discriminated, but their temporal sequence cannot 
be assigned  
correctly (rather, the sequence assignment is more or less at random).
This implies that the discrimination of temporally
distinct events and their sequentialization are different perceptual 
capabilities. 

These results, which were found for different sensory modalities
\cite{poep2,ruhn}, suggest the existence of two different kinds of temporal 
thresholds for the discrimination and sequentialization of perceived events: \\       
(1) a so-called fusion threshold (or transduction threshold) which can be
interpreted as an elementary integration interval for discriminating perceived events.
This threshold is modality-dependent. 
While the mentioned value of approximately 3 msec refers to auditory perception,
the fusion threshold in visual and tactile perception is of the order of 10 msec 
\cite{poep2}. \\
(2) a so-called order threshold of approximately 30 msec which can be interpreted
as an elementary integration interval for the capability to assign sequential order
to perceived events. 
This modality-independent threshold is often characterized as an extended period of
nowness. 

While the fusion threshold can be explained by transduction properties
of signals in the brain, a proper understanding of the order threshold  
remains a topic of vivid discussion. Since its size ($\approx$ 30 msec) 
is the same
for different modalities, it was speculated that the order threshold might be
related to the problem of how pieces of information from an external event, 
which are received in terms of different sensory modalities, are bound together 
such that the external event is perceived 
as a whole (binding problem). In addition, the approximate equivalence 
of $\approx$ 20--40 msec with $\approx$ 30--50 Hz ($\gamma$-band) 
brain activity suggests that the order threshold could be related 
to collective neural oscillations as first reported in \cite{reit,gray}.
For a recent approach to understand the relation between the order threshold
and collective oscillations see \cite{atm2}.

The nature of the order threshold as an elementary integration interval
prevents the sequentialization of successive stimuli with a temporal
interval smaller than approximately 30 msec. This strongly suggests the idea
to use it as a generic lower bound of the time $\Delta T$ between successive 
observations in the quantum Zeno effect. Observations with smaller temporal
distance cannot be perceptually time-ordered. The fact that the order threshold
is modality-independent and its fundamental significance for the binding problem
add to the plausibility of this suggestion.   

\subsection{The significance of $t_0$}

In the quantum Zeno scenario, $t_0$ is the oscillation period of the transition process 
between the considered non-stationary states, under the assumption that there
is no observation and the evolution of the system is governed by $U(t)$. According to
eq.~\ref{allgemein} and setting $C = 1$, observation
leads to an increase of the effective oscillation time from $t_0$ to 
\begin{equation}
\label{empirical}
       T = \frac{t_0^2}{\Delta T}
. 
\end{equation}

With $T \approx 3$ sec and $\Delta T \gtrapprox 30$ msec, this provides 
$t_0 \gtrapprox 300$ msec. Under the influence of observations at a temporal
distance of 30 msec, the observation-free oscillation period of 300 msec
due to $U(t)$ is increased to an oscillation period of 3 sec. It has to be understood, 
though, that the value of $t_0$ is as approximate as those of $T$ and $\Delta T$.
A cognitive time scale corresponding to $t_0$ should, thus, be of the order 
of some hundred milliseconds. 

Roughly speaking, this is the order of magnitude which is most often discussed
as the time required for a stimulus to become ``conscious''. It essentially consists 
of the time required for signal transduction from sensory input to the relevant 
parts of the cortex plus ``unconscious'' preprocessing. Neurophysiological studies  
with event-related potentials show a distinctive universal signature after 
approximately 300 msec, the so-called P300 component, sometimes referring
to additional features up to 900 msec \cite{sutt,donc}. It is independent 
of the specific stimulus and has been attributed to the fact that the perception 
of a stimulus is a process demanding conscious attention (see \cite{korn1} and
references therein). From another perspective, it has been proposed that 
any cognitive processing requires about 100 msec to lead to a consciously
available result \cite{lehm}, e.g.~in terms of a representation.     

The significance of the hundred millisecond time scale with respect to the
conscious availability of a mental representation offers an intuitive understanding 
of $t_0$ in the cognitive context. In contrast to $T$, which represents the ``lifetime'' (or 
mean first passage time) of each of the perceptual representations, $t_0$ can be regarded
as the transition time between those representations. The relaxation
into each one of the representations is much faster than the lifetime in each representation
due to the Zeno effect. Without Zeno effect, the lifetime $T$ would be more or less
identical with the transition time $t_0$.

It has been observed that the lifetime $T$, i.e., the inverse switching rate, for bistable 
Necker cube perception changes considerably if the stimulus is presented in a 
non-continuous way \cite{orba}. Particular combinations of on- and off-time intervals lead to 
a significantly enhanced value of $T$. Recent observations \cite{korn1}
show that $T$ depends essentially on off-times rather than on on-times.
$T$ is maximal for long off-times (on the order of a second).

The relation between $\Delta T$, $T$ and $t_0$ implies that $T$ increases 
if $\Delta T$ is decreased or $t_0$ is increased. Is it possible to interpret 
the situation for non-continuous presentation in such a way that either
$\Delta T$ or $t_0$ is effectively changed due to the discrete on- and
off-times and, thus, leads to an increase of $T$? 

Let us first consider the time scale $\Delta T \approx 30$ msec. 
It represents an intrinsically (due to the operation of the 
cognitive system) given lower bound to the time between observations.
Therefore, it would be implausible to consider significantly smaller values 
of $\Delta T$. Moreover, there is no empirical evidence at hand that 
an increase of $\Delta T$ could be forced by long off-times in non-continuous 
presentation. 

On the other hand, long off-times obviously increase the interval after
which a reversal of the Necker cube perception becomes possible at all. 
>From the theoretical point of view outlined in 
Sec.~2, non-continuous presentation of the Necker cube with considerable off-times
effectively modifies the Hamiltonian of the system, leading to an increased oscillation 
time $t_0$. In detail, this argument applies if off-times are greater than the value of $t_0$
under continuous presentation (with vanishing off-time). For non-continuous presentation, 
off-times (provided they are long enough) can therefore be identified with $t_0$ and 
utilized for an experimentally well-controlled variation of its numerical value. For such
a situation, Fig.~1 shows experimental results 
for $\Delta T = f(t_0)$ from \cite{korn1,orba} together with a theoretically obtained
curve according to eq.~\ref{empirical} and with $\Delta T = 70$ msec. The theoretical 
curve fits the empirical results perfectly well. Using $\Delta T = 70$ msec to estimate
$t_0$ for continuous presentation provides $t_0 \approx 460$ msec. 

\bigskip
\begin{figure}[here]
\begin{center}
\epsfig{figure=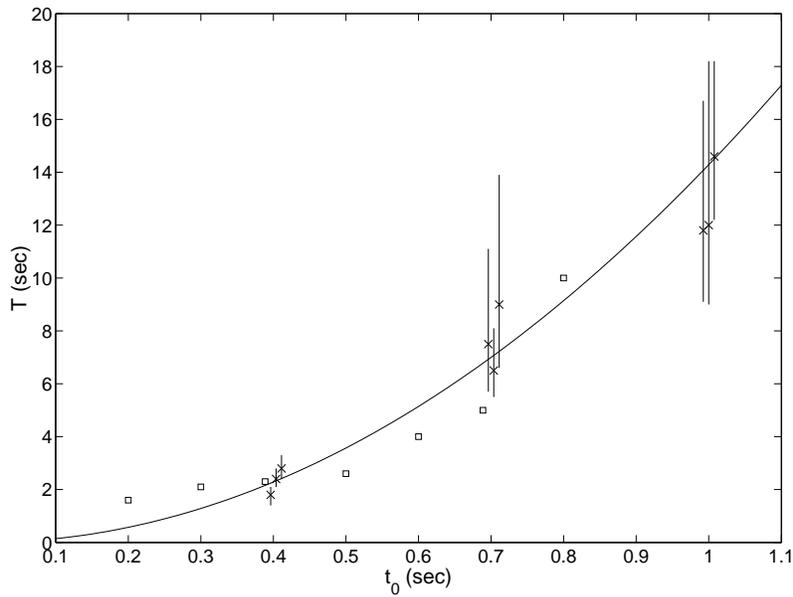,width=300pt}
\end{center}
{\small \caption{Experimentally obtained lifetimes $T$ (inverse switching rates) for the bistable
perception of a non-continuously presented Necker cube.  Crosses mark results from 
Kornmeier \cite{korn1}: for each off-time $t_0$, $T$ (including standard errors) is plotted for 
three on-times of 0.05 sec, 0.1 sec, and 0.4 sec. Squares mark results from Orbach et al.~\cite{orba}
for an on-time of 0.3 sec (no errors indicated in \cite{orba}). The plotted 
curve shows $T$ as a function of off-times $t_0$ according to eq.~\ref{empirical} with
$\Delta T = 70$ msec.} }    
\end{figure}

\section{Summary and conclusions}

In the framework of a generalized, ``weak'' quantum theory, the quantum Zeno effect 
was applied to the bistable perception of ambiguous stimuli. In this application
to a phenomenon of cognitive science, which obviously exceeds the domain of
standard quantum physics, the quantum feature of measurement corresponds
to the act of perception. The main result of this paper is the derivation and
cognitive interpretation of three crucial time scales: 
\begin{description}
\item[$\Delta T$:]
the time between successive observations; as a lower bound for $\Delta T$ in 
cognitive operations, the so-called order threshold of $\approx 30$ msec 
is proposed.
\item[$t_0$:]
the oscillation period for the switching process between two non-stationary states
under the unperturbed evolution of the system; $t_0$ will be increased to $T$
if measurements are carried out.  
\item[$T$:]
the increased switching period for the case that measurements are carried out;
cognitively this corresponds to the switching period of $\approx 3$ sec for bistable 
perception under continuous presentation of the stimulus. 
\end{description}

The quantum Zeno model establishes a relation between these three time scales,
yielding $t_0 \gtrapprox 300$ msec for $\Delta T \gtrapprox 30$ msec and $T \approx 3$ sec
under the condition of continuous stimulus presentation. Such a value of $t_0$ is often 
discussed as the approximate size of the time interval  
after which the processing of sensory inputs leads to a consciously available result,
i.e.~a mental representation of the stimulus. 

Non-continuous presentation of the Necker cube provides the possibility to vary $t_0$
in an experimentally controlled manner in terms of off-times between stimulus
presentation. Empirical results for $T$ as a function of $t_0$ support the relation
between the three times scales with respect to their cognitive significance.
However, further experimental material will be needed to firmly establish the 
cognitive relevance of the quantum Zeno time scales.     

Another prediction of the cognitive quantum Zeno effect is the existence of superpositions
$c_1 \psi_1 + c_2 \psi_2$ of the individual well-defined perception states $\psi_1$ and
$\psi_2$. It is presumably not easy to prepare and observe such superposition states.
Nevertheless, they should appear at least as unstable, transitional states between
$\psi_1$ and $\psi_2$. Elsewhere \cite{atm3} such states were tentatively described as 
``acategoreal''
states, indicating that they do not encode categoreal representations in the usual sense
of cognitive science. In \cite {korn2}, a specific neural correlate of such states was for the 
first time reported in terms of an early (250 msec) component in event-related potentials. 

An interesting option for the preparation of superposition states is the perception
of paradoxical rather than ambiguous figures.  In this case, the key idea is that both 
alternatives of perception operate as repellors rather than attractors, thus pushing the 
state of the system toward the unstable superposition in between them. So far, no 
experimental material is available concerning such a scenario.  

\section*{Acknowledgments}

We are grateful to Werner Ehm, J\"urgen Kornmeier and Jiri Wackermann for helpful 
discussions.

%

\end{document}